# A Revised 27-day Recurrence Index

H. H. Sargent

**Abstract** The original 110 year long 27-day Recurrence Index (original R27) was published more than forty years ago. That index, based on the autocorrelation of consecutive 27-day sets of the geomagnetic aa-index, is a measure of the cycle-to-cycle stability of high speed solar wind structure. During an effort to extend the index, it was discovered that the index could be significantly strengthened by pre-smoothing the geomagnetic aa-index listing used as input. A revised index (revised R27) is presented which clearly shows periods of long term stable solar wind structure toward the end of every sunspot cycle over the last 150 years. The extension of R27 over an interval including the greater part of the space age enables the updating of various studies of long-term solar wind variability based on R27, as well as comparison of R27 with more recently-developed solar-terrestrial parameters.

## 1. Introduction

   In the early Seventies, forecasters at the National Oceanic and Atmospheric Administration's (NOAA) Space Environment Services Center were using a modified form of Bartels chart (Bartels, 1932) to aid in the prediction of geomagnetic activity as reported daily by the U.S. Geological Survey mid-latitude geomagnetic observatory in Fredericksburg, MD. These charts were maintained by hand, and at times were useful forecast tools. If recurrent behavior was seen in the charts, forecasters would usually predict a disturbance of similar intensity and duration (often several days) to follow with the next synodic rotation of the Sun.

   To study when in the solar cycle recurrence could be expected to work as a forecast tool, a 27-day Recurrence Index (R27) (Sargent, 1979) was developed based entirely on Mayaud's (1973) geomagnetic aa-index. R27 was calculated by auto-correlating successive Numbered Bartels Rotations (BR) (Bartels, 1934) of the aa-values, and smoothing the results using a smoothing function similar to that customarily used in smoothing monthly mean sunspot numbers.

R27 clearly showed long periods during the declining phase of each sunspot cycle when recurrent behavior was strongest; particularly during the last ninety years. This was evidence of long term (sometimes as long as five years) stability in the structure of high speed solar wind streams (Sargent, 1985).

By 2019 an additional forty years of aa-index values were available, which overlapped many space-based measurements of solar wind parameters; and an extension of the R27 was planned.

## 2. Methodology

### 2.1 Calculation of the Index

A fresh extended R27 was calculated as follows:

1. All aa-index values were downloaded from the International Service of Geomagnetic Indices (ISGI) over the Internet for the period 1868 to 2020.
2. Daily average values beginning with January 1, 1868 were extracted from the whole data set.
3. A running autocorrelation of each 27-day span of daily values with each succeeding 27-day span was performed and assigned to the date of the first day of the second span. This exercise produced ~152 years of daily correlation coefficients. The correlation coefficients were spot checked with an online linear correlation coefficient calculator app.
4. The overall listing was trimmed to contain only full calendar month's values and a coefficient average was calculated for each month. These monthly average values were scaled by a factor of 100 (only positive whole numbers were of interest).
5. The scaled monthly mean values were smoothed in the same manner as smoothed monthly mean sunspot numbers.

## 2.2 Revision of the Index

The original R27 was calculated using half-day values of the aa-index (fifty four 12-hour averages per rotation) rather than the daily values (twenty seven 24-hour averages per rotation).

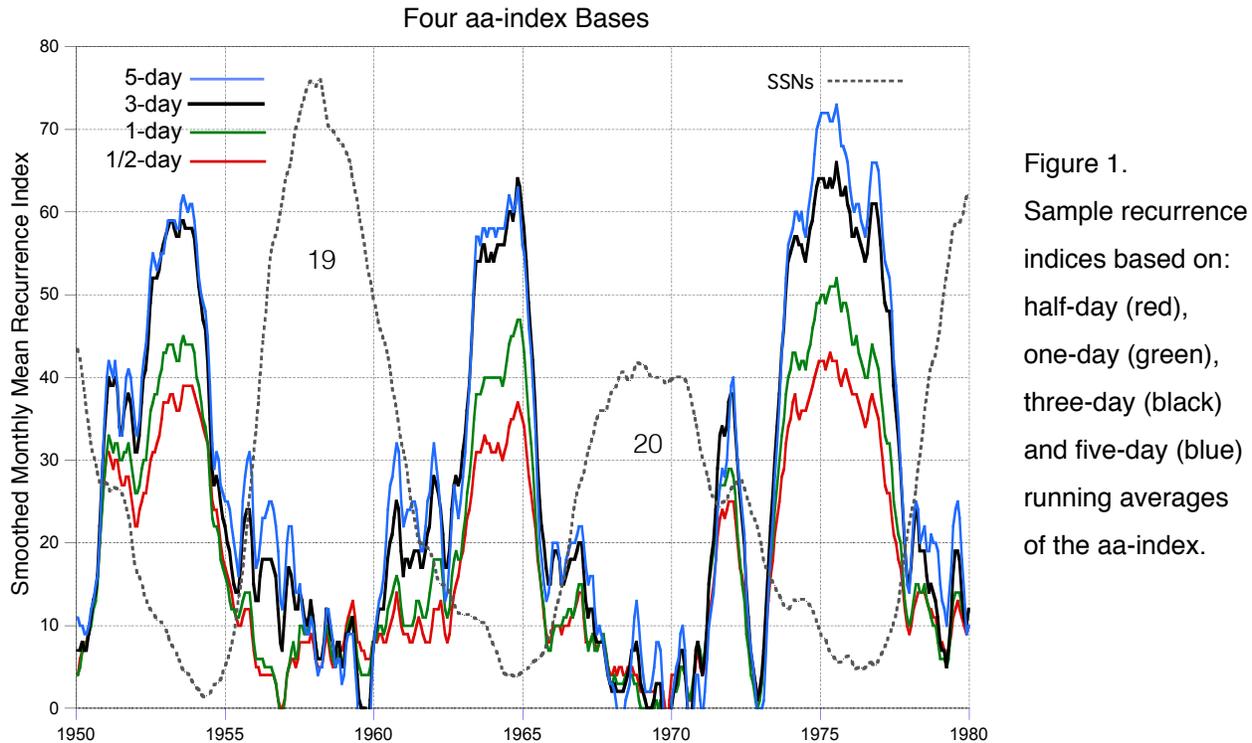

Figure 1. Sample recurrence indices based on: half-day (red), one-day (green), three-day (black) and five-day (blue) running averages of the aa-index.

When the extended R27 was calculated, differences between the original R27 listing and the current R27 listing were found which could only be explained by having used the daily aa-values versus the half day aa-values as input. Consequently, the R27 was recalculated using three-day and five-day running averages of the aa-index as input. A sample of the results is shown in Figure 1.

Clearly, the three- and five-day inputs were significantly stronger. The R27 was thus revised using a three-day running average of the aa-index as input. While the five-day version was slightly stronger at times; it was not used because of concerns about possibly over smoothing the input data.

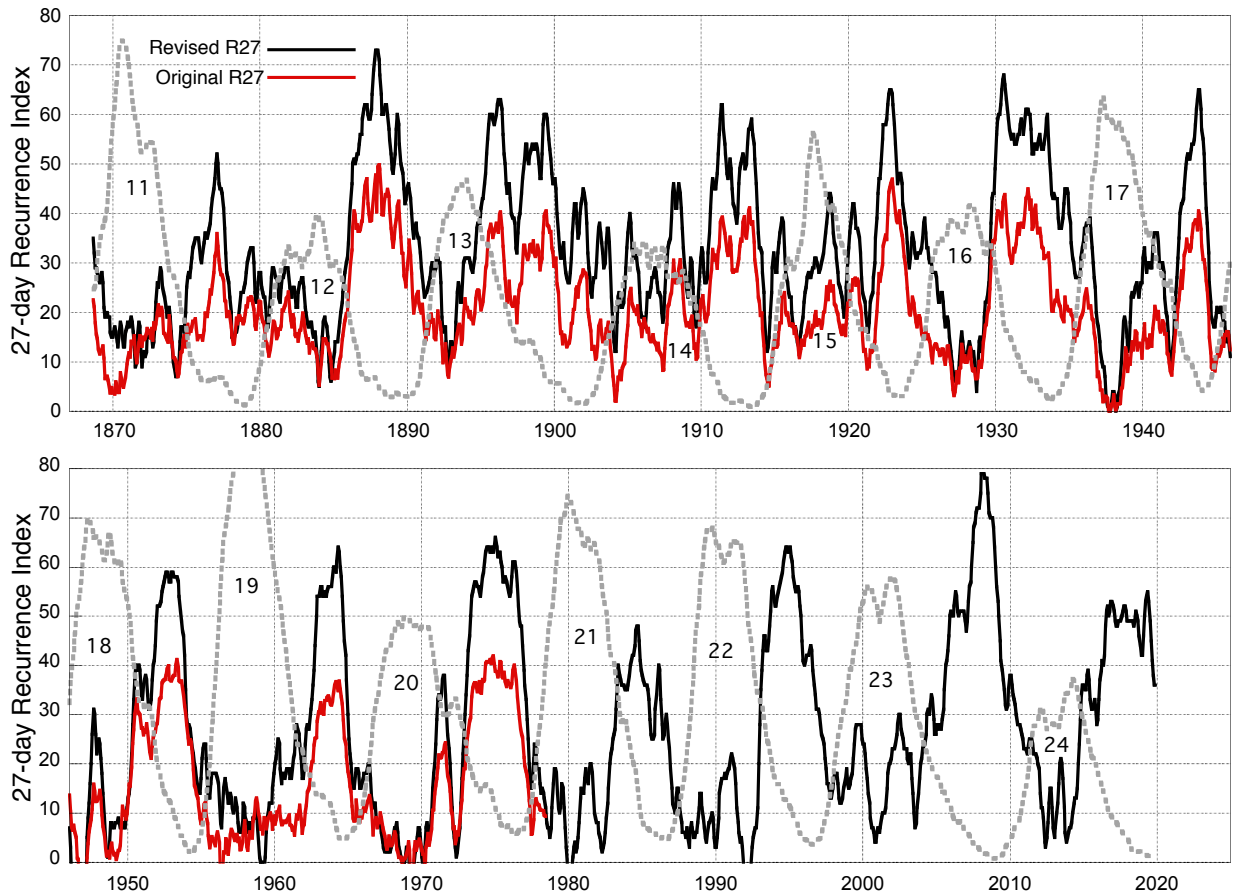

Figure 2. Comparison of Original R27 (1978) and Revised R27 (2020)

## 3. Summary and Discussion

While adding an additional forty years to the original R27 it was found that the Index could be significantly strengthened by preprocessing the input aa-values. Linear correlation coefficients below r = 0.3 are generally considered weak; and coefficients above r = 0.5 are considered moderate to strong. The revision of R27 has clearly significantly strengthened the index. See Figure 2.

In the early days of space weather forecasting, forecasters were looking at the charts for disturbances typically lasting for several days; not for short term spikes that sometimes appeared 27-days apart. Haines, et al (2019) showed it is reasonable to expect storms of two or three days duration.

Now, by applying simple low pass filtering (3-day smoothing) to the input data, the R27 better discriminates against accidental short period disturbances. Thus, the revised R27 further exploits an added data dimension to the 50 year old geomagnetic aa-index, to clearly show long periods toward the end of every sunspot cycle where there is evidently stable structure in high speed solar wind streams.

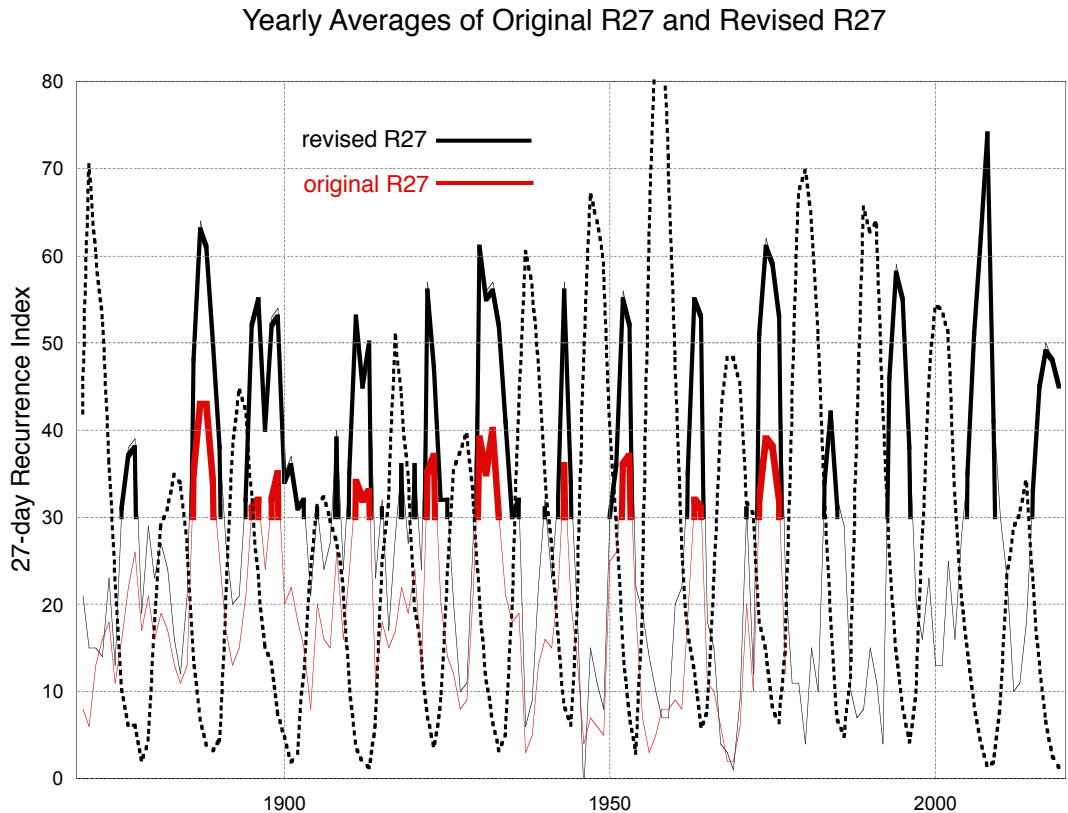

Figure 3. Masked plot of yearly averages of originalR27 (red) and revisedR27 (black) against yearly average sunspot numbers. Values below 30 (linear correlation coefficient of r = 0.3) are considered weak and are masked to simplify plot.

Figure 3 shows the annual average original R27 used in a study twenty years ago by Lockwood, Stamper, and Wild (1999). Future studies of this sort would surely benefit from having the revised R27 available to them. In recent work by McIntosh,, et al (2019), in the early years of every new sunspot cycle, McIntosh's 'terminators' line up perfectly with the precipitous collapse of the revised R27; suggesting possible support of new theoretical work.

**Acknowledgements** This revision of R27, in retirement, would not have been possible without the easy access to up-to-date listing of the aa-index over the Internet from ISGI. Likewise, access to v2.0 (Clette, et al, 2015) of the smoothed monthly mean international sunspot numbers, obtained over the Internet from SILSO (Sunspot Index and Longterm Solar Observations) in Brussels, provided a temporal setting for plots of R27. It is unlikely this work would have been completed without encouragement from Ed Cliver and Scott McIntosh.